\documentclass[pre, amsmath, amssymb, preprintnumbers, showpacs, superscriptaddress, twocolumn, floatfix]{revtex4-1}
\usepackage[pdftex]{graphicx}
\usepackage{amsmath,amssymb,amsfonts,amscd,amsfonts,epsf,epsfig,url}
\usepackage{txfonts}
\usepackage{color}
\usepackage[whole]{bxcjkjatype} 
\usepackage[unicode]{hyperref} 

\begin{document}
\title{Universality in Chaos: Lyapunov Spectrum and Random Matrix Theory }

\author{Masanori Hanada}
\affiliation{Yukawa Institute for Theoretical Physics, Kyoto University, Kitashirakawa Oiwakecho, Sakyo-ku, Kyoto 606-8502, Japan}
\affiliation{Stanford Institute for Theoretical Physics, Stanford University, Stanford, California 94305, USA}
\affiliation{Nuclear and Chemical Sciences Division, Lawrence Livermore National Laboratory, Livermore, California 94550, USA}
\affiliation{Hakubi Center for Advanced Research, Kyoto University, Yoshida Ushinomiyacho, Sakyo-ku, Kyoto 606-8501, Japan}
\author{Hidehiko Shimada}
\affiliation{KEK Theory Center, High Energy Accelerator Research Organization, Tsukuba 305-0801, Japan}
\affiliation{Mathematical and Theoretical Physics Unit, OIST Graduate University, 1919-1 Tancha, Onna-son, Okinawa 904-0495 Japan}
\author{Masaki Tezuka}
\affiliation{Department of Physics, Kyoto University, Kyoto 606-8502, Japan} 

\preprint{YITP-17-17, KEK-TH 1957}

\begin{abstract}
We propose the existence of a new universality in classical chaotic systems 
when the number of degrees of freedom is large: 
the statistical property of the Lyapunov spectrum is described by Random Matrix Theory (RMT).
We demonstrate it by studying the finite-time Lyapunov exponents of the matrix model of a stringy black hole and the mass deformed models. 
The massless limit, which has a dual string theory interpretation, is special 
in that the universal behavior can be seen already at $t=0$, 
while in other cases it sets in at late time.  
The same pattern is demonstrated also in the product of random matrices. 

\end{abstract}

\maketitle

\section{Introduction and Summary}
In this paper we suggest that
the statistical property of the Lyapunov spectrum in classical chaotic systems with a large number of degrees of freedom
is described universally by Random Matrix Theory (RMT).
More precisely, we consider the spectrum of the {\it finite-time} Lyapunov exponents, which is defined 
from the growth of small perturbations during a finite time interval $t$.
Unlike the majority of the previous references in which $t\to\infty$ is taken first,    
we will take the limit of large number of degrees of freedom at each finite $t$ \cite{footnote_limit}. 
This is a natural limit which leads to various universal results such as 
the universal bound on the Lyapunov exponent \cite{Maldacena:2015waa}. 

Our initial motivation was in a different kind of universality in {\it quantum} many-body chaos, 
which has been a hot topic in string theory and quantum information communities in recent years (see e.g. \cite{Sekino:2008he,Maldacena:2015waa}). 
It has been argued that the largest Lyapunov exponent $\lambda_{\rm max}$ has to satisfy a certain bound, 
and the black hole in general relativity saturates the bound \cite{Maldacena:2015waa}. 
In this context G.~Gur-Ari, S.~Shenker and one of the authors (M.~H.) have studied \cite{Gur-Ari:2015rcq} the Lyapunov exponents of a {\it classical} matrix model 
(the D0-brane matrix model) \cite{deWit:1988wri,Witten:1995im,Banks:1996vh,Itzhaki:1998dd}
which is related to a quantum black hole 
with stringy corrections via the gauge/gravity duality \cite{Maldacena:1997re,Itzhaki:1998dd}. They found that the global distribution of the Lyapunov exponents 
follows the semi-circle law near the edge, which is a characteristic feature of the energy spectrum of RMT. 
This suggested the existence of certain universal behaviors in the Lyapunov spectrum of such systems. 

Motivated by this observation, we studied the statistical property of the Lyapunov spectrum in the matrix model \cite{footnote_reference}. 
As we will show, its statistical property is described by RMT for all $t$. 
When we introduce the mass deformation, the RMT description is lost for small $t$.  
However, it does emerge for large $t$.
The spectrum of the product of random matrices, which has been studied as 
an analytically tractable model of chaos, admits the same RMT description.
This is true in other models as well; 
some examples will be reported in \cite{HST_to_appear}.  
Based on these results, we conjecture that the Lyapunov exponents of a large class of many-body chaos, both deterministic and nondeterministic, are described by RMT at late time.

\section{Lyapunov exponent and Lyapunov spectrum}
Let us consider the phase space consisting of $K$ variables, $\phi_i$ ($i=1,2,\cdots,K$). 
By solving the equations of motion, the classical trajectory $\phi_i(t)$ is obtained depending on the initial condition at $t=0$. 
When a small perturbation is added at $t=0$, $\phi_i\to\phi_i+\delta\phi_i$, 
the time evolution of the perturbation can be evaluated by solving the equations of motions with the perturbed initial condition. 
When $\delta\phi_i$ is infinitesimally small, the evolution is described by the transfer matrix $T_{ij}(t,t')$ ($t>t'$)
as $\delta\phi_i(t)=\sum_jT_{ij}(t,t')\delta\phi_{j}(t')$. Let
$a_1(t,t')\ge a_2(t,t')\ge\cdots\ge a_K(t,t')>0$
be the singular values of $T_{ij}(t,t')$. 
The time-dependent Lyapunov exponent $\lambda_i(t,t')$ is defined by $\lambda_i(t,t')=\frac{\log a_i(t,t')}{t-t'}$. 

When the trajectory is bounded, the exponents have unique limits $\lim_{t-t'\to\infty}\lambda_i(t,t')$. 
Usually they are called the Lyapunov exponents. An existence of a positive exponent characterizes 
the sensitivity to the initial condition, which is a necessary condition for the chaos. 

In this paper we consider the finite-time exponents, and study their statistical properties at large $K$. 
Note that we take the large-$K$ limit for each fixed time interval $t-t'$, 
and use many samples which are generated from different initial conditions. 
Two limits, $K\to\infty$ and $t-t'\to\infty$, may or may not commute, depending on the systems \cite{footnote_limit}. 
In chaotic systems, generic initial states evolve to `typical' states after some time, and the statistics is dominated by them. 
We will pick up only typical states.  
It can be achieved by taking $t$ to be sufficiently late time. For the simplicity of the notation, 
we will redefine the time and set $t'=0$,  and call $\lambda_i(t,0)$ as $\lambda_i(t)$. 

In order to compare the statistical property of the Lyapunov spectrum with RMT, 
we use the standard unfolding method \cite{Brody:1981cx}. 
Note that $\{\lambda_i(t)\}$ and $\{a_i(t)\}$ lead to the same unfolded distribution. 
Hence the universality of the Lyapunov exponents discussed in this paper is equivalent to
the universality in the singular values of the transfer matrix describing the linear response.

\section{D0-brane matrix model}
In \cite{Gur-Ari:2015rcq}, the classical limit of the matrix model of 
D0-branes has been considered \cite{footnote_BFSS_reference}. 
The Lagrangian is given by 
\begin{eqnarray}
  L =
  \frac{N}{2}{\rm Tr}\left(
  \sum_{I}(D_tX_I)^2
  +
  \frac{1}{2}
  \sum_{I\neq J}[X_I,X_J]^2
  \right) , 
  \label{Lagrangian}
\end{eqnarray}
where $X_I$ $(I=1,\dots,d)$ are $N\times N$ traceless Hermitian matrices;  
$D_tX_I = \partial_t X_I-[A_t,X_I]$,  where $A_t$ is the $SU(N)$ gauge field.
The number of the traceless Hermitian matrices is $d=9$. 
This system has a scaling symmetry which relates solutions with different energies. 
We will employ a natural energy scale $E=6(N^2-1)-27$ \cite{footnote_normalization},
which corresponds to the unit temperature, $k_{\rm B}T=1$. 
We use the same simulation code as in \cite{Gur-Ari:2015rcq}.

In the $A_t=0$ gauge, the equation of motion is 
\begin{eqnarray}
 \frac{d^2X_I}{dt^2}
 =
\sum_J [X_J,[X_I,X_J]], 
\end{eqnarray}
supplemented with the Gauss's law constraint 
\begin{eqnarray}
\sum_I\left[\frac{dX_I}{dt},X_I\right]=0. 
\end{eqnarray}
By following the procedures explained in \cite{Gur-Ari:2015rcq}, 
we can study the Lyapunov exponents. 
In \cite{Gur-Ari:2015rcq}, it has been observed that the spectrum of $\lambda$ is well approximated by
\begin{eqnarray}
\rho(\lambda,t)
=
\frac{3}{4\tilde{\lambda}_{\rm max}^{3/2}}
\sqrt{\tilde{\lambda}_{\rm max}-|\lambda|}, 
\end{eqnarray}
where $\tilde{\lambda}_{\rm max}$ is a time-dependent parameter which 
approximately equals to the largest Lyapunov exponent. 
Near the edge $|\lambda|\sim\tilde{\lambda}_{\rm max}$, 
this distribution is equivalent to the semi-circle,
$\sqrt{\tilde{\lambda}_{\rm max}^2-\lambda^2}$. 
This is an indication of a possible connection to RMT.

\begin{figure}[htbp]
\includegraphics[width=7.83cm]{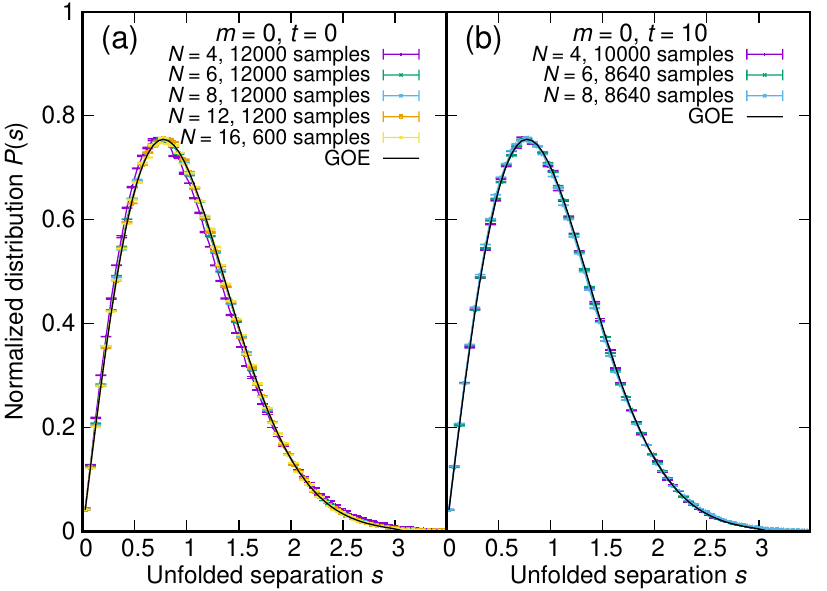}
  \caption{
The separation distribution $P(s)$ for the D0-brane matrix model \eqref{Lagrangian} with $N=4, 6, 8, 12, 16$ at $t=0$ (a), and $N=4, 6, 8$ at $t=10$ (b).
$P(s)$ agrees with $P_{\rm GOE}(s)$ at large $N$. 
  }\label{Fig:N6}
\end{figure}

We have studied the Lyapunov spectrum for $0\leq t \leq 10$ with $N=4, 6, 8$. 
The number of the Lyapunov exponents, which appear in pairs of positive and negative ones with
the same absolute value, is $K=16(N^2-1)$ \cite{footnote_DOF}. 
We ordered the positive exponents as $\lambda_1\ge \lambda_2\ge \cdots$, 
and studied the distribution of the level spacing $s_i\equiv\lambda_{i}-\lambda_{i+1}$. 
From these exponents, the distribution $P(s)$ of the unfolded level separation can be obtained. 
(For the detail of the analysis, including the error estimate, see the supplementary materials.)
It agrees well with the nearest-neighbor level statistics of the GOE ensemble, 
which we denote by $P_{\rm GOE}(s)$ \cite{Dietz:1990}, 
as shown in Fig.~\ref{Fig:N6}, for all values of $t$. 
Already at $t=0$, the spectrum agrees very well with GOE; see Fig.~\ref{Fig:N6} (a). 
Note that we can see a small deviation from GOE at $N=4$. 
Thus the data strongly suggest that 
the level statistics of the finite-time Lyapunov spectrum agrees 
with that of GOE at any $t$, after taking the large-$N$ limit. 

\subsection{Mass deformation}
Next we add the mass term $\Delta L = -\frac{Nm^2}{4}{\rm Tr}\sum_I X_I^2$ to the D0-brane matrix model. 
The physically meaningful parameter is the dimensionless ratio $E/m$. 
Here we fix the energy to be $E=6(N^2-1)-27$ and change $m$. 
In the limit with an infinite mass, or equivalently the zero-energy limit, 
the theory becomes a free theory, which is not chaotic \cite{footnote_mass}. 

\begin{figure}[htbp]
\includegraphics[width=7.83cm]{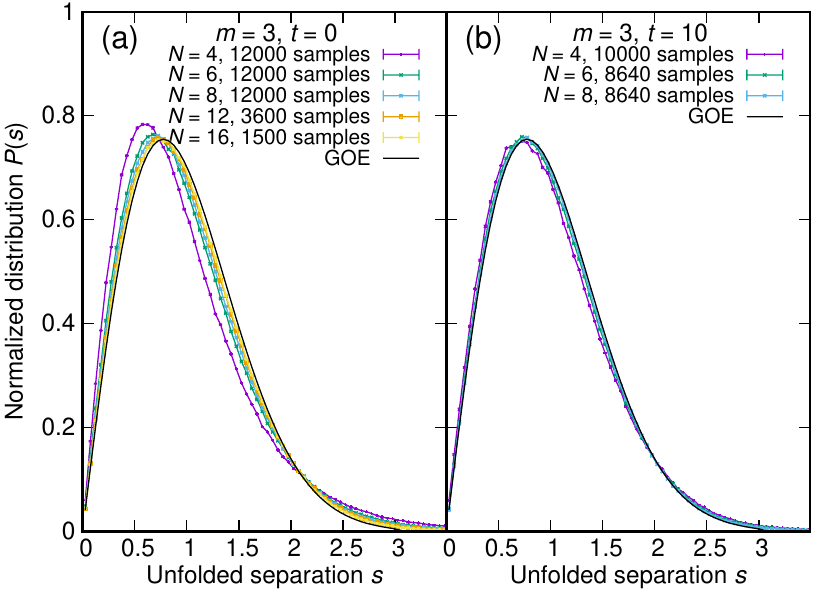}
  \caption{
The separation distribution $P(s)$ for the D0-brane matrix model \eqref{Lagrangian} with the mass deformation, $m=3$, $N=4, 6, 8, 12, 16$ at $t=0$ (a), and $N=4, 6, 8$ at $t=10$ (b).
At $m\neq 0$, although $P(s)$ and $P_{\rm GOE}(s)$ do not agree at $t=0$, 
they become very close at $t=10$. 
  }\label{Fig:BFSSMass-sep}
\end{figure}

In Fig.~\ref{Fig:BFSSMass-sep} (a) the distribution of the unfolded level separations
with $m=3$ is shown. Although it is linear in $s$ for small $s$, indicating level repulsion between Lyapunov exponents,  
the distribution disagrees with that of GOE, having a peak at smaller $s$ 
and a longer tail.
However, as shown in Fig.~\ref{Fig:BFSSMass-sep} (b), the distribution goes close to GOE at
$t>0$. 

\begin{figure}[htbp]
\includegraphics[width=7.83cm]{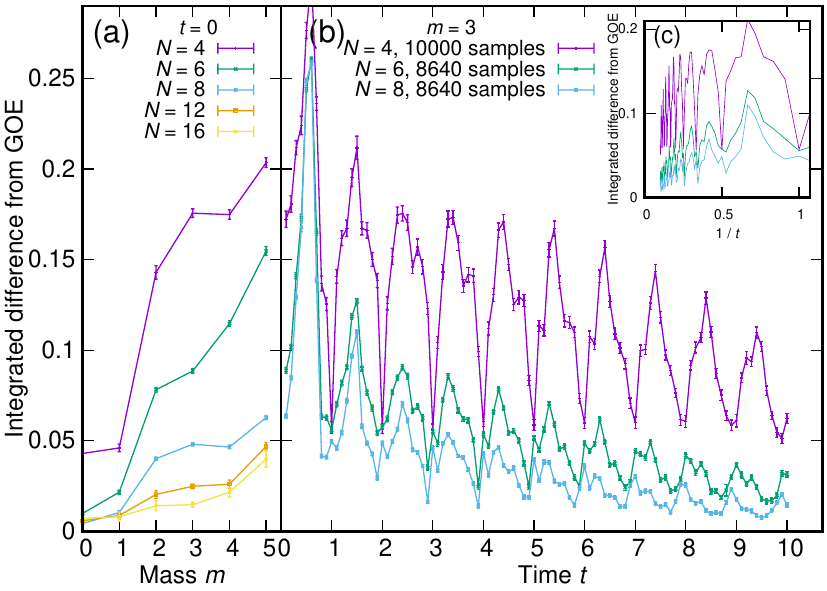}
  \caption{
(a): Mass dependence of the difference between the mass-deformed model and the GOE random matrix, $\int ds|P(s)-P_{\rm GOE}(s)|$.
The sample size is 12000 for $N=4, 6, 8$ and at least 1000 (230) for $N=12$ ($16$), respectively.
(b): Time dependence of the difference, $m=3$, 
$N=4,6,8$, with the same quantity plotted against $1/t$ in the inset (c).
The difference oscillates and gradually decreases. 
At $N=8$, the decreases at late time is $\sim 1/t$. 
  }\label{Fig:N6_m0-3}
\end{figure}

To make this observation more precise we calculated the difference,
$\int ds|P(\lambda)-P_{\rm GOE}(\lambda)|$, 
of the distribution from that of GOE.
The difference is plotted at $t=0$ for several values of $m$ in 
Fig.~\ref{Fig:N6_m0-3} (a). The spectrum disagrees with that of GOE at finite $m$, 
and the deviation is larger when $m$ is larger. 
In Fig.~\ref{Fig:N6_m0-3} (b), the time dependence is shown for $m=3$, $N=4,6,8$.
The deviation from $P_{\rm GOE}(s)$ oscillates, and gradually decreases.  
This result strongly suggests that the distribution converges to 
$P_{\rm GOE}(s)$  when the limit $t\to\infty$ is taken after $N\to\infty$.

\subsection{Beyond nearest neighbor}
In order to see the agreement with RMT beyond the nearest-neighbor level correlation, 
we have compared the spectral form factor (SFF) defined by
\begin{eqnarray}
Z(\tau)=\sum_n e^{i\lambda_n\tau} 
\end{eqnarray}
and its RMT counterpart for Gaussian symmetric random matrices of the same dimension $K$,
\begin{eqnarray}
Z_{\mathrm{GOE}(K)}(\tau)=\sum_n e^{iE_n\tau}.  
\end{eqnarray}
The spectral form factor captures more information about the spectrum, the so-called spectral rigidity. 
The large $\tau$ behavior of the SFF reflects the fine grained structure of the energy spectrum. 
The small $\tau$ region is sensitive to the global shape of the spectrum, which is not expected to be universal.

\begin{figure}[htbp]
\includegraphics[width=7.83cm]{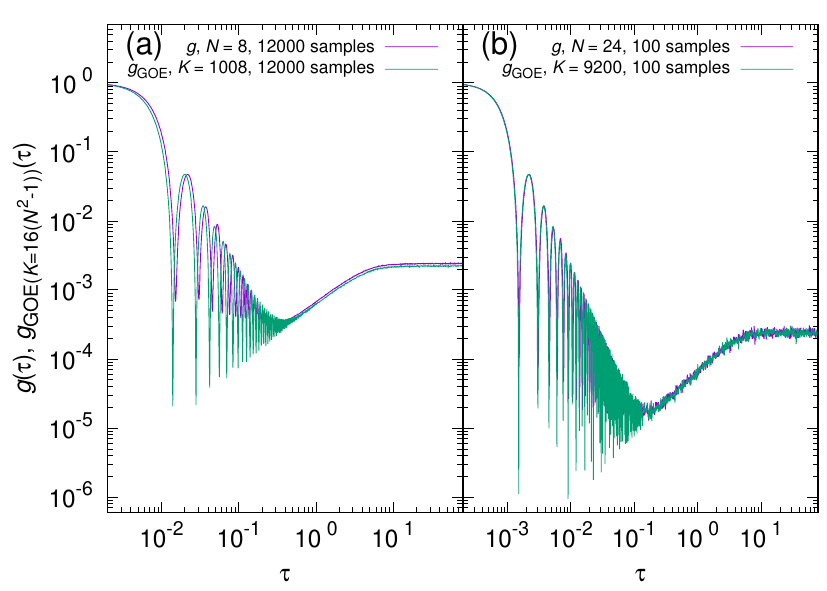}
  \caption{
The SFF $g(\tau)$, at $\beta = 0$ for the unfolded Lyapunov spectrum of the D0-brane matrix model \eqref{Lagrangian} with $N=8$ (left) and $N=24$ (right) at $t=0$ and for the unfolded eigenvalues of Gaussian random symmetric matrices with dimension $K=16(N^2-1)$.
}\label{Fig:N20-24}
\end{figure}

In Fig.~\ref{Fig:N20-24} we have plotted $g(\tau) \equiv |Z(\tau)|^2 / K^2$ calculated from the Lyapunov spectrum of 
the BFSS matrix model at $t=0$ and $g_{\mathrm{GOE}(K)}(\tau) \equiv |Z_{\mathrm{GOE}(K)}(\tau)|^2 / K^2$. 
The agreement at large $\tau$ (the ramp $\sim\tau^1$ and the plateau $\sim\tau^0$)
means the agreement of the Lyapunov spectrum and RMT energy spectrum beyond the nearest neighbor. 
Note that the disagreement in the small $\tau$ region is not a problem, it simply means the global shapes of the spectrum are different. 

\begin{figure}[htbp]
\includegraphics[width=7.83cm]{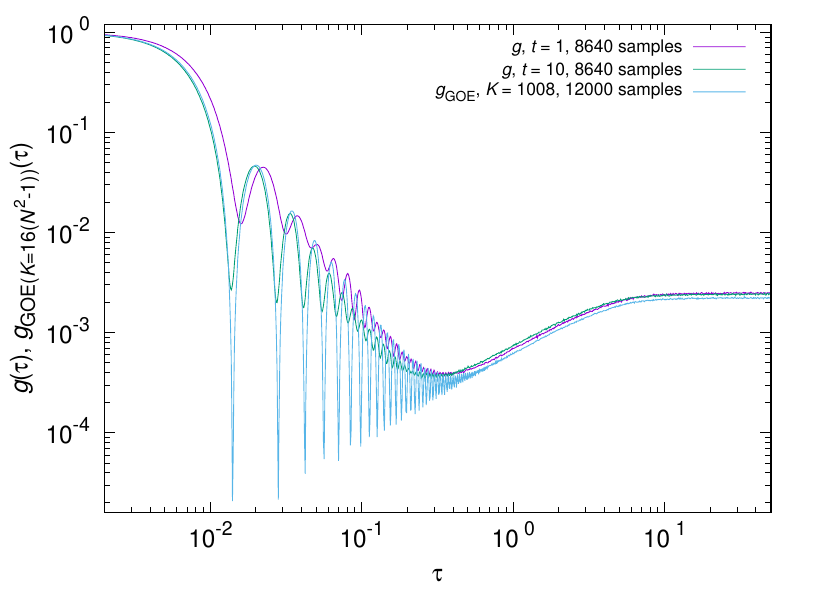}
\caption{
The SFF $g(\tau)$ for the unfolded Lyapunov spectra of the mass-deformed model with $N=8$ and $m=3$ for $t=1$ and $10$, 
and for the unfolded Gaussian random symmetric matrix eigenvalues with $K=16(N^2-1)=1008$.
}\label{Fig:SFFgtN8m3}
\end{figure}

We repeated the same analysis with a mass deformation. In Fig.~\ref{Fig:SFFgtN8m3}, 
the SFFs $g(\tau)$ for the mass-deformed model with $N=8$ and $m=3$ for $t=1$ and $t=10$ are shown. 
The convergence to RMT at late time (large $t$) can be seen very clearly. 

\section{Product of random matrices}
Let us consider a product of $t$
matrices randomly chosen from a certain ensemble
(`Random Matrix Product', RMP), 
\begin{eqnarray} 
{\cal M}(t)
=
M_tM_{t-1}\cdots M_2M_1. 
\end{eqnarray}
We take the matrix size to be $K\times K$. 
The RMP has been studied as a toy model of the Lyapunov growth, by regarding $M_i$ to be 
an analogue of the transfer matrix at a short time separation. 
From the singular values $a_i(t) (i=1,2,\cdots,K)$, ordered as $a_1(t) \geq a_2(t) \geq \cdots \geq a_K(t)$, 
we define the finite-time Lyapunov exponents by $\lambda_i(t)=(\log a_i(t))/t$. 

The RMP has also been considered in the study of quantum transport phenomena,
such as the conduction of electrons in a disordered wire~\cite{RefQT}.
Our analysis in this section is closely related to results in the literature of the
quantum transport phenomena; our $K$ corresponds 
to the number of transport channels, and
$t$ corresponds to the length of the disordered wire~\cite{footnote_QTvsGOE}.
In quantum transport phenomena, the evolution is studied of the transmission eigenvalues
when the length of the wire is changed~\cite{EvolutionQuantumTransport}.
It would be interesting to consider the time evolution of Lyapunov spectrums 
of the classical (deterministic or non-deterministic) chaotic systems 
from a similar point of view.

If each $M_i$ is a real matrix (also a complex matrix) 
with the weight $e^{-K{\rm Tr}MM^\dagger}$, then the level spacing statics of Lyapunov exponents $\lambda_i(t)$
follow that of the standard GOE (GUE) 
for any fixed $t$.
This is easily verified numerically,
and for the complex matrices an analytic derivation can be found in  
\cite{ProductRandomMatrices}. 
This is precisely analogous with the case of the massless D0-brane matrix model
\eqref{Lagrangian}. 
Note that $t\to\infty$ with fixed $K$ is different from RMT \cite{Newman1986}\cite{footnote_larget}.

\begin{figure}[htbp]
\includegraphics[width=7.83cm]{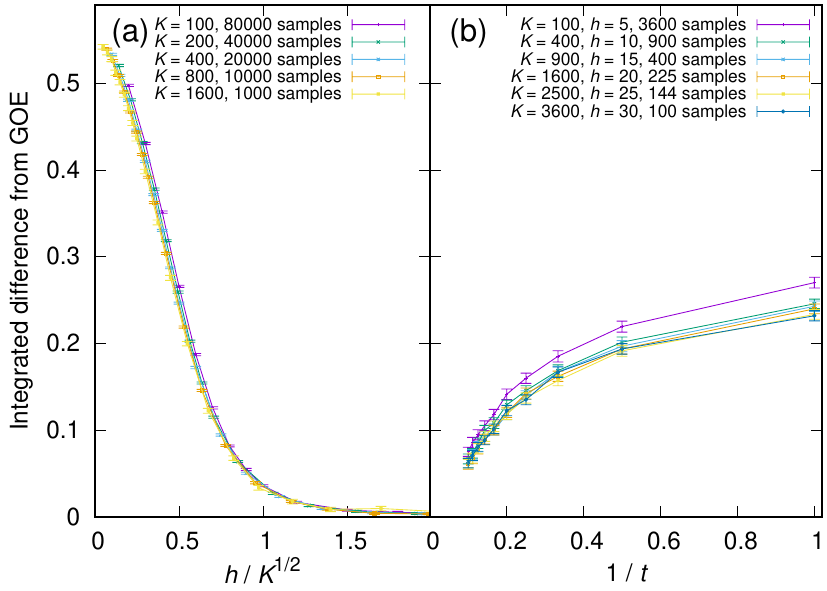}
\caption{
(a): The difference from GOE, $\int ds|P(s)-P_{\rm GOE}(s)|$, at $t=1$, as a function of $h/\sqrt{K}$. We can see that 
the difference converges to an $O(1)$ value when $h/\sqrt{K}$ is fixed. 
(b): The same quantity for various $K$ and $t$, with $h/\sqrt{K}=1/2$.
A clear convergence to GOE at large $K$ and large $t$ can be seen. 
}
\label{Fig:RMP-GOE}
\end{figure}

One can also introduce a deformation of the RMP playing a role
analogous to the mass deformation of the matrix model.
We have numerically studied a product of 
real-valued random band matrices, whose $(i,j)$ components are set to zero unless $\vert i-j\vert < h$, with the periodic identification $i\sim i+K$. 
As shown in Fig.~\ref{Fig:RMP-GOE} (a), 
the deviation of $P(s)$ from GOE at $t=1$ 
converges to an $O(K^0)$ value in the large-$K$ limit 
when $h/\sqrt{K}$ is fixed. In Fig.~\ref{Fig:RMP-GOE} (b),
the results for the products with $h/\sqrt{K}=1/2$ are shown.  
At large $t$, the plot shows a clear tendency of the convergence to GOE.

\begin{figure}[htbp]
\includegraphics[width=7.83cm]{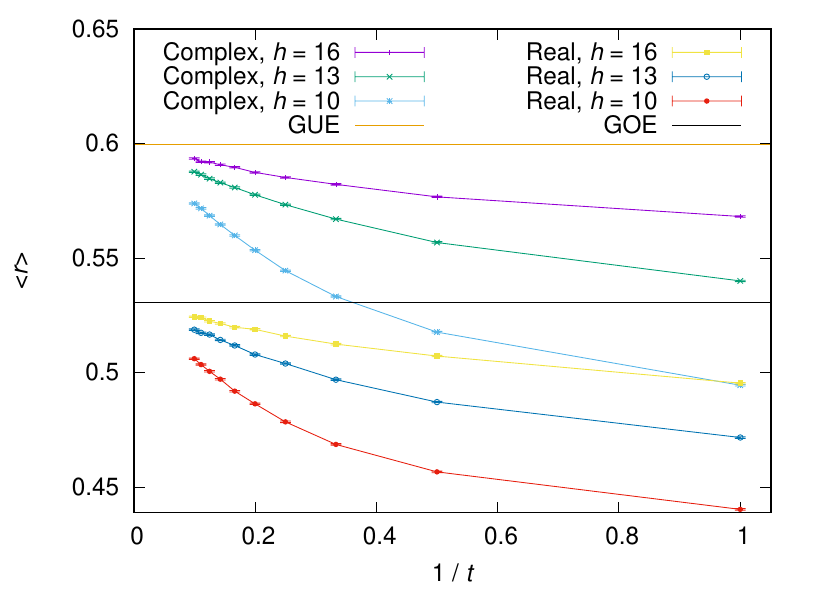}
\caption{
The average nearest neighbor gap ratio $\langle r \rangle$ plotted against the inverse of the number of multiplied matrices, $1/t$, for the complex and real random matrix products with $K=900$ and $h=16, 13, 10$. The sample size is $1000$ for all cases.
The values for GUE and GOE random matrix eigenvalues from \cite{Atas2013} are also shown by horizontal lines for comparison.}
\label{Fig:gapRatio}
\end{figure}

We also calculate the average nearest neighbor gap, defined by
\begin{equation}
\langle r\rangle = \left\langle \frac{\min(s_{i}, s_{i+1})}{\max(s_{i}, s_{i+1})} \right\rangle_i,
\label{Eqn:nngap}
\end{equation}
in which $s_{i} = \lambda_{i} - \lambda_{i+1}$ and the average $\langle \cdots \rangle$ is taken over $i=1,\ldots, K-2$ and all the samples. The average nearest neighbor gap characterizes the correlation between the neighboring gaps in the spectrum.
In Fig.~\ref{Fig:gapRatio} we have plotted the value of $\langle r\rangle$, 
both for products of real and complex matrices,  
against the inverse of the number of multiplied matrices $t$, both for complex and real matrices with $K=900$ and $h=16, 13, 10$,
along with the values for GOE and GUE matrices presented in \cite{Atas2013}.
This is the evidence that the universality holds for next-to-next nearest neighboring levels.

\begin{figure}[htbp]
\includegraphics[width=7.83cm]{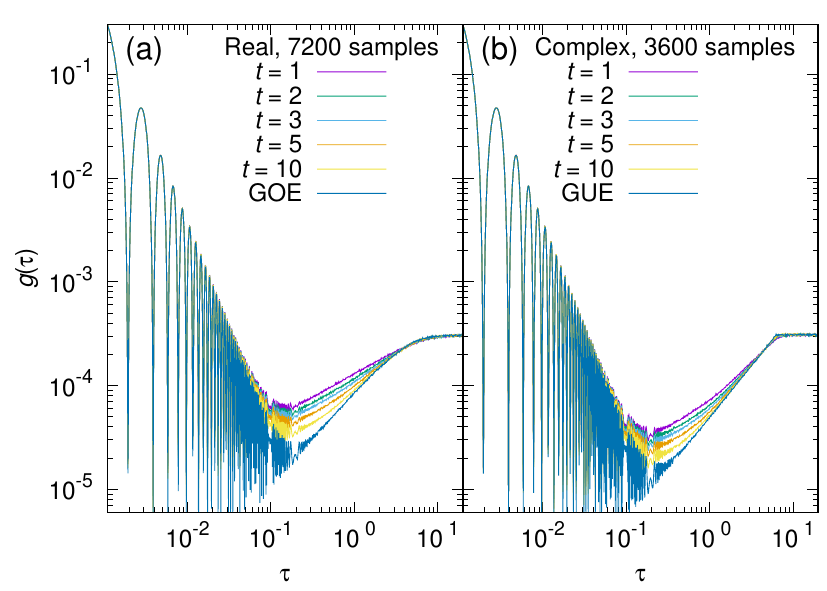}
\caption{
    (a) ((b)): $g(\tau) = \vert Z(\tau)\vert^2/\vert Z(\tau=0)\vert^2$ for the finite-time Lyapunov exponents obtained from the singular values of $t$ real (complex) random matrix products with $K=3600$ and $h=32$, compared against $g_\mathrm{GOE (GUE)}(\tau) = \vert Z_\mathrm{GOE (GUE)}(\tau)\vert^2/\vert Z_\mathrm{GOE (GUE)}(\tau=0)\vert^2$ obtained from GOE (GUE) random matrices of the same dimension $K$. We have used the unfolded spectrum. 
 See \cite{RMP-unfold-detail} for the detail of the unfolding.
}\label{Fig:gtGOE-GUE}
 \end{figure}

Furthermore, in order to see the correlation over even larger separations, in Fig.~\ref{Fig:gtGOE-GUE} (a)
we have compared the SFFs for the product of real matrices, $|Z(\tau)|^2/|Z(\tau=0)|^2$, with that of GOE random matrices, $|Z_{\rm GOE}(\tau)|^2/|Z_{\rm GOE}(\tau=0)|^2$. 
We can see that $|Z(\tau)|^2$ approaches to $|Z_{\rm GOE}(\tau)|^2$ as $t$ increases.  
Also in Fig.~\ref{Fig:gtGOE-GUE} (b) we have plotted $g(\tau)$ for complex random matrix products against $g_\mathrm{GUE}(\tau)$ obtained from GUE random matrices.
Here again, we can see the agreement between the finite-time Lyapunov exponents and RMT energy spectrum beyond the nearest neighbors.

\section{Discussions}
In this paper we have suggested the existence of a 
new universality in the Lyapunov spectrum of the classical chaotic systems
based on numerical evidence for the matrix models and random matrix products. 
The massless D0-brane matrix model and the product of un-banded 
Gaussian random matrices are special 
in that the universal behavior can be seen at any time scale. 
It is interesting to speculate that
other Yang-Mills theories and/or quantum gravitational systems 
satisfy the same property. Classical field theory calculations 
which are useful for this direction can be found in e.g. \cite{Bolte:1999th,Kunihiro:2010tg}. 

We have also studied several other systems, e.g. 3d Coulomb gas, coupled Lorenz attractors 
and coupled logistic maps, and observed qualitative evidence for the same universality \cite{HST_to_appear}. In general, the scaling of $t$ and the number of degrees of freedom
 should be carefully studied. 
For example, although the random matrix product with fixed $h$ and fixed $t$ does not become RMT, 
it is likely that $h$ fixed and $t\sim K^p$, with a certain power $p>0$, can lead to RMT. 

A possible path toward an understanding of the mechanism behind the universality is 
to see how the spectra of various systems converge to RMT. 
As we commented in section IV, the classical chaotic systems 
and quantum transport phenomena are mathematically closely related, and thus
it may be possible to deepen understanding of existence of universalities by considering
both phenomena together.
It may also provide us with a new characterization of various chaotic systems;
the amount of deviation from RMT may be reflecting the strength of chaos, 
and the special property in the D0-brane matrix model would be related 
to the fast scrambling \cite{Sekino:2008he,Maldacena:2015waa}. 
The generalization of this universality to the quantum chaos would be even more interesting. 
We hope that the study of the statistical properties of 
the Lyapunov exponents provides us with a new viewpoint for studying chaotic systems.

{\bf \textit{Acknowledgement:}}
We would like to thank S.~Aoki, P.~Buividovich, P.~Damgaard, E.~Dyer, A.~M.~Garc\'ia-Garc\'ia, G.~Gur-Ari, S.~Hikami, J.~Magan, S.~Nishigaki, S.~Sasa,
A.~Sch\"{a}fer, S.~Shenker, A.~Streicher, K.~Takeuchi, A.~Ueda, P.~Vranas and M.~Walter for discussions.

This work was partially supported by JSPS KAKENHI Grant Numbers JP25287046 (M.H.), JP17K14285 (M.H.), 
JP15H05855 (M.T.), JP26870284 (M.T.), JP17K17822 (M.T.) and JP16H06490 (H.S.).
Part of computation in this work was performed at Supercomputer Center,
Institute for Solid State Physics, University of Tokyo.


\section*{Supplementary materials}

\subsection{Details of the analysis of the unfolded spectrum:}

We explain how we produced the plots in this paper.
We take $W$ independent samples labelled by $w=1,2,\ldots,W$.
Each sample consists of $K$ Lyapunov exponents 
$\lambda_1^{(w)} \ge \lambda_2^{(w)}\ge\ldots\ge\lambda_K^{(w)}$.

We first make a histogram with bins of width $\Delta\lambda$ using all $W$ samples.
There are
$WK$ exponents in total.
We then normalize the histogram so that $\int \rho(\lambda) d\lambda=\sum_i \rho_i \Delta\lambda = 1$, where $i$ is a label for the bins. 
For $\mathcal{O}(10^7)$ exponents we use in the majority of our plots, we typically take $\mathcal{O}(10^3)$ bins.

For Hamiltonian systems discussed in this paper,
all exponents are paired with the exponent of the same absolute value and the opposite sign. 
Therefore we focus on positive Lyapunov exponents.
We further omit both largest $5\%$ and smallest $5\%$ 
of the positive exponents, in order to avoid
the exponents close to the edge affecting the fit discussed below.
We denote the maximum and minimum of retained exponents by $\lambda^{(\mathrm{max})},\lambda^{(\mathrm{min})}$ respectively.
For the bins containing retained exponents we fit the density of exponents $\rho(\lambda)$, by a polynomial
$\tilde{\rho}(\lambda) = \sum_{k=0}^{k_\mathrm{max}} a_k(\lambda-\lambda_0)^k$ of $\lambda$, for unfolding the spectrum.
We typically choose $k_\mathrm{max} = 10$.
To reduce numerical error, $\lambda_0$ is chosen within
the fitting range $[\lambda^{(\mathrm{min})},\lambda^{(\mathrm{max})}]$.

Then the spectrum is `unfolded' by considering
$s_j^{(w)} \equiv S
(\tilde{R}(\lambda_j^{(w)}) - \tilde{R}(\lambda_{j+1}^{(w)}))$,
in which $\tilde{R}(\lambda) = \int_{\lambda_0}^\lambda \tilde{\rho}(\lambda')d\lambda'=\sum_{k=0}^{k_\mathrm{max}}\frac{a_k}{k+1}(\lambda-\lambda_0)^{k+1}$
and
$S\sim K$ is the normalizing factor chosen so that the average of $s_j^{(w)}$ is unity.

We plot the histogram of $s_j^{(w)}$. Namely, for each bin $[q \Delta s, (q+1) \Delta s)$,
we count the number $n_q$ of $s_j^{(w)}$ within this bin, and take $P(s_q \equiv (q+\frac12)\Delta s) = n_q / (\Delta s \sum_q n_q)$.

From the distribution $P(K,t)$ with given $(K,t)$, 
we define the deviation from the GOE distribution by 
\begin{align}
\Delta (K,t)
&\equiv
\int ds\  \left|P_{K,t}(s)-P_{\rm GOE}(s)\right| \nonumber\\ 
&\simeq \sum_{q=0}^{q_\mathrm{max}} \vert P(s_q) - P_{\mathrm{GOE},q} \vert \Delta s,
\label{def:discrepancy}
\end{align}
in which we have defined $P_{\mathrm{GOE},q}\equiv P_\mathrm{GOE}(s_q)$.

When the average separation is normalized to be 1, the GOE distribution is often approximated by
Wigner's surmise,
\begin{eqnarray}
P_{\rm GOE (Wigner)}(s)
=
\frac{\pi s}{2} e^{-\frac{\pi}{4}s^2}.
\end{eqnarray}
However, for our purpose the Wigner's surmise is not accurate enough.
The correct distribution 
$P_{\rm GOE}(s)$ admits a Taylor series expansion and
a Pad\'e approximant, which are available in \cite{Dietz:1990}.
In our analysis, it is sufficient to use the Taylor series expansion of
$P_{\rm GOE}(s)$ as its approximation for $s\leq 3$.
We use  the upper limit,
$s_{q_\mathrm{max}} \simeq 3$, in the summation (\ref{def:discrepancy}).

\subsection{Error estimate}
Firstly we separate the samples to $L$ groups. We used $L=4$. 
We prepare $L$ data sets, by excluding one of the $L$ groups. 
By using a certain bin size, we make a histogram for each data set, 
and determine the heights $P^{(l)}_{q}$, where $l=1,2,\cdots,L$ is the label for the data set, 
and $q$ is the label for the bin. 
The Jack-knife error is defined by
\begin{align}
\delta P_q\equiv \sqrt{(L-1)\left(\frac{1}{L}\sum_{l=1}^L\left(P^{(l)}_{q}\right)^2 - P_q^2\right)}. 
\end{align}
This error estimate is used for the error-bars in figures \ref{Fig:N6} and \ref{Fig:BFSSMass-sep}.

Let $P_q^{\rm max}\equiv P_q+\delta P_q$ and $P_q^{\rm min}\equiv P_q-\delta P_q$. 
We denote the bin width by $\epsilon$.
We estimate the error-bar for 
$\Delta(K,t)$, which we denote by $\delta^{(\pm)}\left(\Delta(K,t)\right)$, as
\begin{align}
\Delta(K,t) \pm \delta^{(\pm)}\left(\Delta(K,t)\right)%
=
\sum_q
\delta^{(\pm)}\left(\Delta(K,t)\right)_q \Delta s,
\end{align}
where
\begin{align}
&\delta^{(+)}\left(\Delta(K,t)\right)_q\nonumber\\
=&
{\rm max}\left\{
\left|P_q^{\rm max}-P_{{\rm GOE},q}\right|,  
\left|P_q^{\rm min}-P_{{\rm GOE},q}\right| 
\right\},
\end{align}
and
$\delta^{(-)}\left(\Delta(K,t)\right)_q=0$
if $P_i$ and $P_{\rm GOE}$ coincides within the error estimate explained above
({\it i.\ e. } if 
$P_q^{\rm min}\le
P_{{\rm GOE}.q}
\le
P_q^{\rm max}$), otherwise
\begin{align}
&\delta^{(-)}\left(\Delta(K,t)\right)_q\nonumber\\
=&
{\rm min}\left\{
\left|P_q^{\rm max}-P_{{\rm GOE},q}\right|,  
\left|P_q^{\rm min}-P_{{\rm GOE},q}\right| 
\right\}.
\end{align}

\subsection{The Lyapunov spectrum for the D0-brane matrix model}

\begin{figure}[htbp]
\includegraphics[width=7.83cm]{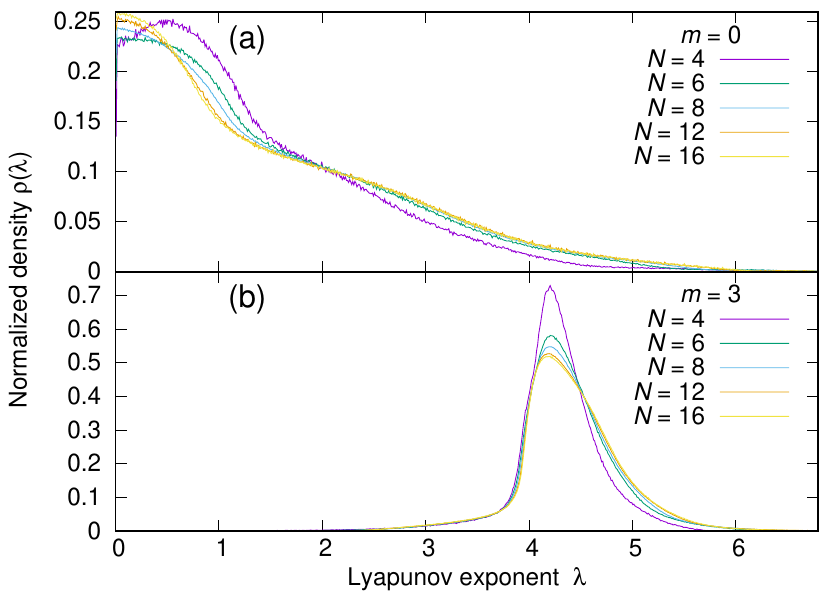}
\caption{The histogram $\rho(\lambda)$ of the local ($t=0$) Lyapunov exponents ($\lambda > 0$)
for the D0-brane matrix model with $m=0$ (a) and $3$ (b), $N=4, 6, 8, 12, 16$.
The bin width is $\Delta \lambda = 0.01$.
The same set of data is used for the left panels of Fig.~\ref{Fig:N6} and \ref{Fig:BFSSMass-sep}.
}
\label{fig:suppl-rho-t0}
\end{figure}
\begin{figure}[htbp]
\includegraphics[width=7.83cm]{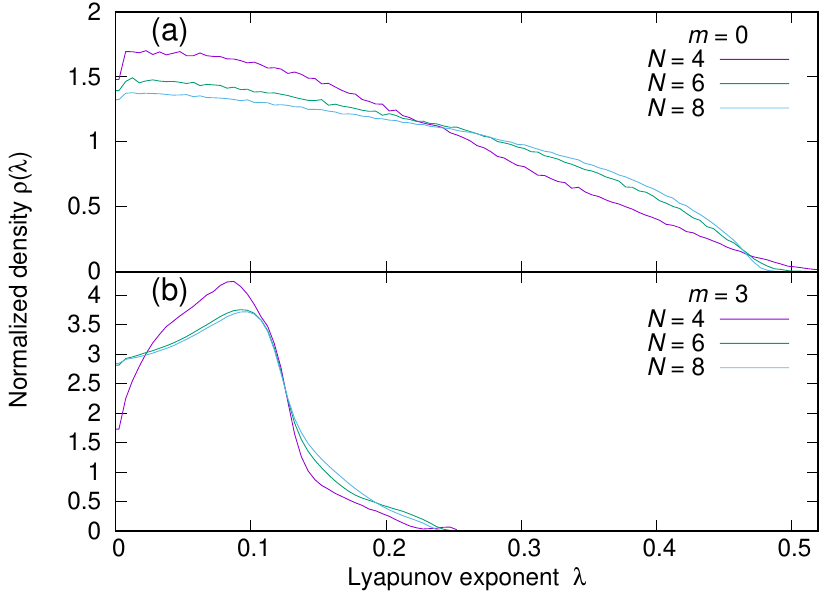}
\caption{The histogram $\rho(\lambda)$ of the Lyapunov exponents
for the D0-brane matrix model at $t=10$ for $m=0$ (a) and $3$ (b), $N=4, 6, 8$.
The bin width is $\Delta \lambda = 0.005$.
The same set of data is used for the right panels of Fig.~\ref{Fig:N6} and \ref{Fig:BFSSMass-sep}.
}
\label{fig:suppl-rho-t10}
\end{figure}

In Figures~\ref{fig:suppl-rho-t0} and \ref{fig:suppl-rho-t10} we plot the Lyapunov spectrum
obtained for the D0-brane matrix model at $t=0$ and $t=10$, respectively.
The plots are symmetric about $\lambda = 0$, therefore we have plotted only the
positive exponents. The data suggest that $\rho(\lambda)$ rapidly approaches the large-$N$ limit.


\begin{thebibliography}{3}

\bibitem{footnote_limit}
This is the limit for the models discussed in this paper. As for other models, 
more generic double scaling of $t$ and the number of degrees of freedom might be needed. 
See {\it Discussions} for the detail. 

\bibitem{Maldacena:2015waa} 
  J.~Maldacena, S.~H.~Shenker and D.~Stanford,
  JHEP {\bf 1608}, 106 (2016).




\bibitem{Sekino:2008he} 
  Y.~Sekino and L.~Susskind,
  JHEP {\bf 0810}, 065 (2008).

\bibitem{Gur-Ari:2015rcq} 
  G.~Gur-Ari, M.~Hanada and S.~H.~Shenker,
  JHEP {\bf 1602}, 091 (2016).

\bibitem{deWit:1988wri} 
  B.~de Wit, J.~Hoppe and H.~Nicolai,
  Nucl.\ Phys.\ B {\bf 305}, 545 (1988).


\bibitem{Witten:1995im} 
  E.~Witten,
  Nucl.\ Phys.\ B {\bf 460}, 335 (1996).


\bibitem{Banks:1996vh} 
  T.~Banks, W.~Fischler, S.~H.~Shenker and L.~Susskind,
  \prd\ {\bf 55}, 5112 (1997).


\bibitem{Itzhaki:1998dd} 
  N.~Itzhaki, J.~M.~Maldacena, J.~Sonnenschein and S.~Yankielowicz,
  \prd\ {\bf 58}, 046004 (1998),


\bibitem{Maldacena:1997re} 
  J.~M.~Maldacena,
  Int.\ J.\ Theor.\ Phys.\  {\bf 38}, 1113 (1999)
  [Adv.\ Theor.\ Math.\ Phys.\  {\bf 2}, 231 (1998)].

\bibitem{footnote_reference}
Similar numerical experiments have been performed to a certain disorder system in \cite{Ahlers2001} and \cite{Patra2016}.
There are two important differences from our work: they studied non-chaotic parameter region of the theory
(more precisely, one of the parameter choice in \cite{Patra2016} is chaotic due to the $1/N$-correction), 
and they have considered different limit from ours: $t\to\infty$ for each fixed system size. 
Note that, in their case, statistical analysis can be performed by varying the disorder parameters.
Interestingly, the latter observed a reasonable agreement with RMT in certain parameter regions.

\bibitem{Ahlers2001}
V.~Ahlers, R.~Zillmer, and A.~Pikovsky, 
\pre\ {\bf 63}, 036213 (2001).

\bibitem{Patra2016}
S.~K.~Patra and A.~Ghosh, 
\pre\ {\bf 93}, 032208 (2016).


\bibitem{HST_to_appear}
M.~Hanada, H.~Shimada and M.~Tezuka, in progress. 



\bibitem{Brody:1981cx} 
  T.~A.~Brody, J.~Flores, J.~B.~French, P.~A.~Mello, A.~Pandey and S.~S.~M.~Wong,
  \rmp\  {\bf 53}, 385 (1981).



\bibitem{footnote_BFSS_reference}
Previous studies of the same system include \cite{Matinyan:1981dj,Savvidy:1982wx,Savvidy:1982jk,Asplund:2011qj,Asplund:2012tg,Asano:2015eha,Aoki:2015uha}.
The nature of chaos has been explored in \cite{Savvidy:1982wx,Savvidy:1982jk,Aref'eva:1997es,Aref'eva:1998mk,Asplund:2011qj,Asplund:2012tg,Asano:2015eha,Aoki:2015uha}.
In particular, \cite{Asplund:2011qj,Asplund:2012tg,Aoki:2015uha} studied the decay in time of two-point functions, 
and \cite{Aref'eva:1997es} studied the Lyapunov behavior.


\bibitem{Matinyan:1981dj} 
  S.~G.~Matinyan, G.~K.~Savvidy and N.~G.~Ter-Arutunian Savvidy,
  Sov.\ Phys.\ JETP {\bf 53}, 421 (1981)
  [Zh.\ Eksp.\ Teor.\ Fiz.\  {\bf 80}, 830 (1981)].


\bibitem{Savvidy:1982wx}
  G.~K.~Savvidy,
  Phys.\ Lett.\ B {\bf 130} (1983) 303.


\bibitem{Savvidy:1982jk} 
  G.~K.~Savvidy,
  Nucl.\ Phys.\ B {\bf 246}, 302 (1984).


\bibitem{Aref'eva:1997es} 
  I.~Y.~Aref'eva, P.~B.~Medvedev, O.~A.~Rytchkov and I.~V.~Volovich,
  Chaos Solitons Fractals {\bf 10}, 213 (1999).




\bibitem{Aref'eva:1998mk} 
  I.~Y.~Aref'eva, A.~S.~Koshelev and P.~B.~Medvedev,
  Mod.\ Phys.\ Lett.\ A {\bf 13}, 2481 (1998).

\bibitem{Asplund:2011qj} 
  C.~Asplund, D.~Berenstein and D.~Trancanelli,
  \prl\ {\bf 107}, 171602 (2011).

\bibitem{Asplund:2012tg} 
  C.~T.~Asplund, D.~Berenstein and E.~Dzienkowski,
  \prd\ {\bf 87}, 084044 (2013).

\bibitem{Asano:2015eha} 
  Y.~Asano, D.~Kawai and K.~Yoshida,
  JHEP {\bf 1506}, 191 (2015).

\bibitem{Aoki:2015uha} 
  S.~Aoki, M.~Hanada and N.~Iizuka,
  JHEP {\bf 1507}, 029 (2015).

\bibitem{footnote_normalization}
We keep the energy per degree of freedom to be fixed when $K$ is sent to infinity. 
This is the 't Hooft large-$N$ limit. 

\bibitem{footnote_DOF}
There are $18(N^2-1)$ components corresponding to $X_I$ and $\frac{dX_I}{dt}$. 
The Gauss's law constraint eliminates $N^2-1$ of them, and 
the residual gauge symmetry removes another $N^2-1$.  


\bibitem{Dietz:1990}
For calculation of $P_{\rm GOE}(s)$, we have followed 
B.~Dietz and F.~Haake,
Z.\ Phys.\ B {\bf 80}, 153 (1990).


\bibitem{footnote_mass}
It is free at $m\to\infty$ and finite $t$. If $t\to\infty$ is taken first, it is still chaotic.

\bibitem{RefQT}
Earlier contributions include 
B.~L.~Al'tshuler, and B.~I.~Shklovski\u{\i}, Zh. Eksp. Teor. Fiz. {\bf 91}, 220 (1986) [Sov. Phys. JETP 64, 127 (1986)];
Y.~Imry, EPL (Europhysics Letters) {\bf 1}, 249 (1986);
J.~-L.~Pichard, and G.~Sarma, Journal of Physics C {\bf 14}, L127 (1981);
K.~A.~Muttalib, J.~-L.~Pichard, and A.~Douglas~Stone, Phys. Rev. Lett. {\bf 59}, 2475 (1987);
J.~-L.~Pichard, N.~Zanon, Y.~Imry, and A.~Douglas~Stone, J. Phys. France {\bf 51}, 587 (1990).
For reviews, see 
A.~Douglas~Stone, P.~A.~Mello, K.~A.~Muttalib, and J.~-L.~Pichard in Mesoscopic Phenomena in Solids, ed. by B.~L.~Altshuler, P.~A.~Lee, and R.~A.~Webb (North-Holland, Amsterdam 1991), p. 369; 
C.~W.~J.~Beenakker, Rev. Mod. Phys. {\bf 69}, 731 (1997).
We wish to thank an anonymous reviewer for pointing out to us
the relevance of and for a careful explanation of results in the quantum transport phenomena.

\bibitem{footnote_QTvsGOE}
It is known that some quantities associated with the transmission eigenvalues
studied in the quantum transport phenomena do not agree with those for the GOE. 
See reviews cited in \cite{RefQT}.
For the observables we consider in this paper, {\it i. e.} 
the unfolded level-spacing distribution 
and the spectral form factor, we see agreement between the Lyapunov spectrum of 
classical chaotic systems and the GOE. In spite of this,
it is quite possible that for other quantities associated with finer details of the spectrum, 
the Lyapunov spectrum may show agreement with the ensembles 
studied in the quantum transport phenomena rather than the GOE.

\bibitem{EvolutionQuantumTransport}
O.~N.~Dorokhov, ZhETF Pis. Red {\bf 36}, 259 (1982) [JETP Lett. {\bf 36}, 318 (1982)];
O.~N.~Dorokhov, Zh. Eksp. Teor. Fiz. {\bf 85}, 1040 (1983) [Sov. Phys. JETP {\bf 58}, 606 (1983)];
P.~A.~Mello, P.~Pereyra, and N.~Kumar, Ann. Phys. (N.Y.) {\bf 181}, 290 (1988); 
S.~Iida, H.~A.~Weidenm\"uller, and J.~A.~Zuk, Phys. Rev. Lett. {\bf 64}, 583 (1990);
S.~Iida, H.~A.~Weidenm\"uller, and J.~A.~Zuk, Ann. Phys. (N.Y.) {\bf 200}, 219 (1990).


\bibitem{ProductRandomMatrices}
D.-Z.~Liu, D.~Wang, and L.~Zhang, Ann. Inst. H. Poincaré Probab. Statist. {\bf 52}, 1734 (2016).
Ann. Inst. H. Poincaré Probab. Statist. {\bf 52}, 1734 (2016).
For a review, see,  G.~Akemann and J.~R.~Ipsen, Acta Phys. Pol. B {\bf 46}, 1747 (2015). 
See also J.~ R.~Ipsen, H.~Schomerus, J. Phys. A {\bf 49}, 385201 (2016), 
for a continuum time analogue. 
	
	
\bibitem{Newman1986}
C.~M.~Newman,  
Communications in Mathematical Physics, {\bf 103}, 121-126 (1986).

\bibitem{footnote_larget}
As $t$ goes to infinity with fixed $K$, 
the distribution of finite-time exponents becomes infinitely 
concentrated around the exponents defined for the infinite time.
Hence, for example, the unfolded level spacing distribution is very sharply concentrated
around $1$ for large $t$ with fixed $K$, which is clearly different from the distribution
obtained from the GOE.
This phenomena is known as the ``crystallization'' of transmission eigenvalues in the 
context of quantum transport phenomena. 
See reviews cited in \cite{RefQT}.


\bibitem{Atas2013}
Y.~Y.~Atas, E.~Bogomolny, O.~Giraud, and G.~Roux,
\prl\ {\bf 110}, 084101 (2013).


\bibitem{RMP-unfold-detail} 
 We unfolded the central $90~\%$ of the exponents $\{\lambda_j\}_{j=181}^{3420}$ from each sample, using a tenth order polynomial fit of the spectrum with the top $2.5~\%$ and the bottom $2.5~\%$ excluded.


\bibitem{Bolte:1999th} 
  J.~Bolte, B.~Muller and A.~Schafer,
  \prd\ {\bf 61}, 054506 (2000). 


\bibitem{Kunihiro:2010tg} 
  T.~Kunihiro, B.~Muller, A.~Ohnishi, A.~Schafer, T.~T.~Takahashi and A.~Yamamoto,
  \prd\ {\bf 82}, 114015 (2010). 
  	
\end{thebibliography}
\end{document}